\pdfoutput=1 

\documentclass[a4paper, conference]{IEEEtran}

\usepackage[numbers]{natbib}
\usepackage[shortcuts,acronym]{glossaries}
\makenoidxglossaries
\newacronym{dpd}{DPD}{Double Pulsed Direct Method}
\newacronym{rssi}{RSSI}{Received Signal Strength Indication}
\newacronym{tof}{ToF}{Time of Flight}
\newacronym{ftm}{FTM}{Fine Time Measurement}
\newacronym{toa}{ToA}{Time of Arrival}
\newacronym{tdoa}{TDoA}{Time Difference of Arrival}
\newacronym{twr}{TWR}{Two Way Ranging}
\newacronym{sds-twr}{SDS-TWR}{Symmetrical Double Sided Two Way Ranging}
\newacronym{ds-twr}{DS-TWR}{Double Sided Two Way Ranging}
\newacronym{asym-ds-twr}{asym-DS-TWR}{Asymetric Double Sided Two Way Ranging}
\newacronym{djkm}{DJKM}{Djaja-Josko-Kolakowski-Method}
\newacronym{dpw}{DPW}{Double Pulsed Whistle}
\newacronym{mle}{MLE}{Maximum-Likelihood-Estimator}
\newacronym{dpp}{DPP}{Double Pulsed Positioning}
\newacronym{los}{LoS}{Line of Sight}
\newacronym{nlos}{NLoS}{Non Line of Sight}
\newacronym{uwb}{UWB}{Ultra Wide Band}
\newacronym{ppm}{ppm}{Parts per Million}

\usepackage{amsmath}

\usepackage{siunitx}
\usepackage{blindtext}
\usepackage{float}
\usepackage{cleveref}

\ifCLASSINFOpdf
  \usepackage[pdftex]{graphicx}
  \DeclareGraphicsExtensions{.pdf,.jpeg,.png}
\else
  \usepackage[dvips]{graphicx}
  \DeclareGraphicsExtensions{.pdf}
\fi

\ifCLASSOPTIONcompsoc
  \usepackage[caption=false,font=normalsize,labelfont=sf,textfont=sf]{subfig}
\else
  \usepackage[caption=false,font=footnotesize]{subfig}
\fi

\makeatletter
\let\old@ps@headings\ps@headings
\let\old@ps@IEEEtitlepagestyle\ps@IEEEtitlepagestyle
\def\confheader#1{%
\def\ps@headings{%
\old@ps@headings%
\def\@oddhead{\strut\hfill#1\hfill\strut}%
\def\@evenhead{\strut\hfill#1\hfill\strut}%
}%
\def\ps@IEEEtitlepagestyle{%
\old@ps@IEEEtitlepagestyle%
\def\@oddhead{\strut\hfill#1\hfill\strut}%
\def\@evenhead{\strut\hfill#1\hfill\strut}%
}%
\ps@headings%
}
\makeatother

\confheader{2019 International Conference on Indoor Positioning and Indoor Navigation (IPIN), 30 September - 3 October 2019, Pisa, Italy}

\title{A Generalized TDoA/ToA Model for ToF Positioning}

\IEEEoverridecommandlockouts
\IEEEpubid{\makebox[\columnwidth]
       {\hfill 978-1-7281-1788-1/19/\$31.00 ~\copyright~2019 IEEE}
\hspace{\columnsep}\makebox[\columnwidth]{ }}

\begin{document}

\twocolumn[
\centerline{\parbox[c][10cm][c]{10cm}{%
	\selectfont
	{\textbf{\LARGE{Copyright Notice}}\newline\newline
	\textcopyright 2019 IEEE. Personal use of this material is permitted. Permission from IEEE must be obtained for all
	other uses, in any current or future media, including reprinting/republishing this material for advertising
	or promotional purposes, creating new collective works, for resale or redistribution to servers or lists, or
	reuse of any copyrighted component of this work in other works.\newline\newline
	\textbf{Published in: Proceedings of the 2019 International Conference on Indoor Positioning and Indoor Navigation (IPIN), 30 Sept. - 3 Oct. 2019, Pisa, Italy}
	\newline
	\textbf{DOI: 10.1109/IPIN.2019.8911742}
}}}]
\newpage

\author{\IEEEauthorblockN{Maximilian von Tschirschnitz\IEEEauthorrefmark{1},
Marcel Wagner\IEEEauthorrefmark{2}, Marc-Oliver Pahl\IEEEauthorrefmark{3} and Georg Carle\IEEEauthorrefmark{4}}
\IEEEauthorblockA{Technical University Munich,
Intel Deutschland GmbH\\
Email: \IEEEauthorrefmark{1}maximilian.tschirschnitz@tum.de,
\IEEEauthorrefmark{2}marcel.wagner@intel.com,
\IEEEauthorrefmark{3}pahl@s2o.net.in.tum.de,
\IEEEauthorrefmark{4}carle@net.in.tum.de}}


%


\setlength{\topskip}{5pt}

\maketitle
\vspace*{-0.8cm}

\begin{abstract}
Many applications require positioning.
Time of Flight (ToF) methods calculate distances by measuring 
the propagation time of signals.
We present a novel ToF localization method.
Our new approach works infrastructure-less, without pre-defined roles like 
Anchors or Tags.
It generalizes existing synchronization-less Time Difference of Arrival (TDoA) 
and Time of Arrival (ToA) algorithms. 
We show how known algorithms can be derived from our new method. 
A major advantage of our approach is that it provides a comparable or 
better clock error robustness, i.e. the typical errors of crystal oscillators 
have negligible impact for TDoA and ToA measurements. 
We show that our channel usage is for most cases superior compared to the 
state-of-the art. 
\end{abstract}


%
\IEEEpeerreviewmaketitle

\section{Introduction}\label{sct:Introduction}

	The term \gls{tof} refers to a class of methods which calculate distances by measuring propagation time of signals. When measuring \gls{tof} there are two main approaches to determine positions, \gls{toa} and \gls{tdoa}.
	The most relevant approaches include \gls{sds-twr} \cite{hach_ss_twr}, \gls{asym-ds-twr} \cite{alternative_twr}, Whistle \cite{Whistle},
	\gls{dpw} \cite{dpw} and the method published by \citet{anti_whistle} which we will refer to as \gls{djkm}.

	All those methods are developed for specific setups.
	For example, some methods assume that the devices which are located are only capable to send signals (Whistle, \gls{dpw}) or, vice versa, can only receive data (\gls{djkm}). Another important difference is that some methods have been developed for sound signals (e.g. Whistle) or radio signals (e.g. \gls{dpw}, \gls{djkm}). The methods have all in common that no clock synchronization between devices is needed.

	We have conducted a detailed analysis of those methods in a different publication \cite{analysis_paper}.
	While this paper is independent and self contained we recommend the lecture of that other work as an overview of the state-of-the-art of \gls{tof} methods. 

	One major drawback of the aforementioned methods is that they all need clearly defined roles of devices, e.g. there have to be \emph{Anchors} with known positions which locate so called \emph{Tags}. This restricts the ability to support \emph{infrastructureless}-positioning, i.e. positioning without pre-defined roles or with changing roles.
	This makes it impossible for existing \gls{tdoa} methods to determine relative positions between Nodes in scenarios where all devices are mobile.

	Those scenarios include swarms of search and rescue drones, indoor navigation for firefighters and autonomous vehicle coordination.  
	
	This work is presenting a new method which can be seen as a generalization of the aforementioned methods in the sense that all existing methods can be derived from the proposed scheme.
	Our method, called \gls{dpp}, can simultaneously perform \gls{toa} and \gls{tdoa}.
	Even without any clock synchronization the errors induced by clockdrift  are negligible.
        We show that our method outperforms the existing methods in terms
	of accuracy, flexibility and channel-utilization in almost all situations.


        The rest of the paper is structured as follows. We introduce the new method and show how the other methods can be derived from it (Section \ref{sct:dpp_definition}). The clock errors are analyzed and compared with other methods in Section \ref{sct:dpp_error_analysis} followed by the evaluation of \gls{dpp} in Section \ref{sct:dpp_advantages}. The work is concluded with a conclusion and future work. 

\section{DPP System Overview}\label{sct:dpp_definition}
	For the following definitions, we assume that any access to the \gls{tof} signal channel is coordinated by side communication channels (e.g. Wifi), and thus collisions
	(through simultaneous transmissions) are avoided.
	Also we assume that the Nodes can exchange information over a side channel.

	We consider the devices' clocks as imperfect with a typical clock drift accuracy of $\pm20$ppm.
	That means, we can not assume the devices to be synchronized nor to be running on one exact frequency.
	The only assumption we make over the device clocks is that their drift remains constant during one measurement cycle.

	\subsection{Formal Definitions}\label{sct:formal_definitions}
		As outlined in Section \ref{sct:Introduction}, we are not differentiating between Anchors, Tags, Mirrors or such.
		Instead, all devices are represented as \textbf{Nodes} where each Node element $X$ is an element from the set of all Nodes $X \subseteq N$.
		We define a function $P_t(X)$ where $X$ is a Node and $t$ is a timestamp, that returns the cartesian coordinate of the Node $X$ at time $t$.
		As a simplification, we write $P(X)$ to refer to the current position of a Node.
		Further, we define that a \textbf{System} $S$ of Nodes is a subset of the available Nodes ($S \subseteq N$).
		
		A Node is further classified as one of three types:
		\begin{itemize}
			\item{\textit{Passive} Nodes: $N_P$\newline
					Nodes which are only capable of receiving data and  never transmit something}
			\item{\textit{Active} Nodes $N_A$\newline
					Nodes which are only capable of transmitting data and never receive something}
			\item{\textit{Bilateral} Nodes $N_B$\newline
					Nodes which are capable of receiving and transmitting data}
		\end{itemize}
		This differentiation makes up a partitioning of:
		$N$: $$N = N_P \dot\cup N_A \dot\cup N_B$$
		We define $v$ to be the speed of the signal used when transmitting between Nodes (e.g., speed of light).
		The \gls{tof} between Nodes is defined as the euclidean distance between them divided through the signal velocity:
		$$\forall X,Y \in N: d_{XY} := \frac{||P(X) - P(Y)||}{v}$$
		It follows that:
		$$d_{XY} = d_{YX}$$
		The \gls{tdoa} between two Nodes $X,Z$ in relation to signal source $Y$ is, under the previous definition of Nodes, defined as: 
		$$\forall X, Y, Z \in (N_P \cup N_A \cup N_B):$$
		\begin{equation}
			T_{XZ}^Y := d_{YZ} - d_{XY} \label{eqn:dpp_tdoa_definition}
		\end{equation}

		In this document, representative Systems are sometimes selected as examples to visualize certain positioning scenarios.
		When Node Systems are presented in a figure, we use the following symbols:
		\begin{itemize}
			\item Passive Nodes are labeled as $P$ and indexed individually by their suffix
			\item This counts equivalently for Active ($A$) and Bilateral Nodes ($B$)
			\item Nodes with a known position (e.g., infrastructure mounted or Anchor devices) are depicted with a circle surrounding them
		\end{itemize}

	\subsection{DPP Operation Procedure and Message Diagram} \label{sct:operation_dpp}
\begin{figure}
			\centering
			\includegraphics{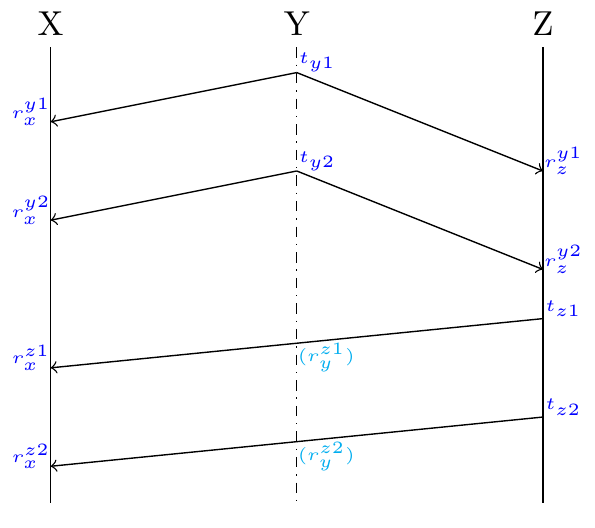}
			\caption{Double Pulsed Positioning (general message diagram extract)}
			\label{fig:dp_pos}
		\end{figure}

		The general operation-procedure of a \gls{dpp}-Node-System is the following:
		\begin{itemize}
			\item{All of the Nodes in $N_A \cup N_B$ sequentially (turn-based) transmit two messages (Double Pulse)}
			\item{All Nodes in $N_P \cup N_B$ listen constantly for any incoming messages}
		\end{itemize}
		When each Active and Bilateral Node has communicated at least once, a so-called \textbf{Cycle} is completed.
		Fig.~\ref{fig:dp_pos} shows an extract from such a sequence, and this way describes the different types of gathered measurements.
		Each receiving timestamp is labeled as $r_x^{yp}$, where $x$ is the Node receiving, $y$ the Node transmitting, and $p$ is the 
		index of the signal of this combination of receiver and sender in the current cycle.
		Similarly, the transmission timestamps are labeled as $t_{yp}$, where $y$ is the transmitting Node and $p$ is the index of that Node's transmission in the current
		cycle with $p$ typically in the range of $p \in \{1, 2\}$.
		
	\subsection{TDoA Value}
		In the following, we derive a formula to calculate \gls{tdoa} values based on the values generated by the \gls{dpp} operation model (\ref{sct:operation_dpp}).

		When comparing the schematic Figures in this paper with \gls{dpw} \cite{dpw}, we can see that some subsets of the message diagram resemble \gls{dpw} schemes.
		That allows us to apply the tools provided by \gls{dpw} to our measurements.

\begin{figure}
			\centering
			\subfloat[$Y$ pulsing and $Z$ ``mirroring'']
			{
				\includegraphics[width=.45\linewidth]{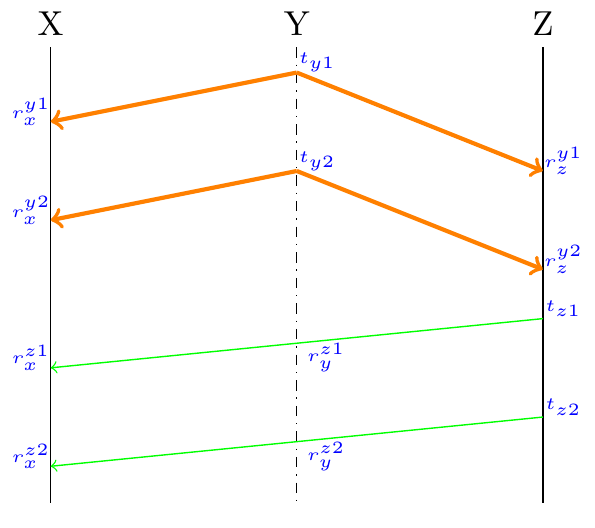}
			}
			\hfill
			\subfloat[$Z$ pulsing and $Y$ ``mirroring'']
			{
				\includegraphics[width=.45\linewidth]{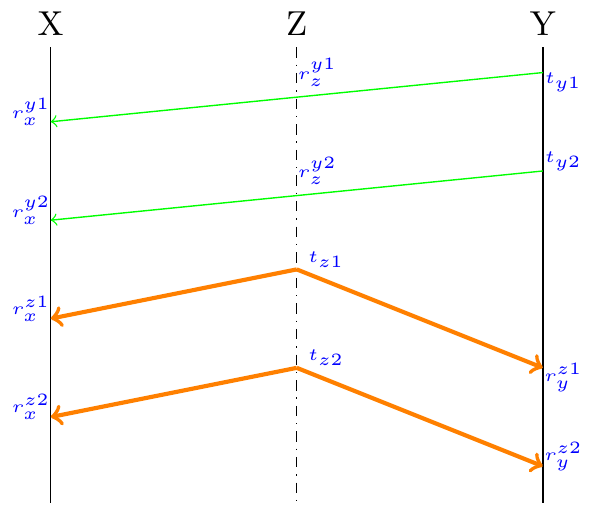}
			}
			\caption{\gls{dpw} sub-schemes in \gls{dpp} protocol extract example (pulse: orange/thick, mirror: green/regular)}
			\label{fig:dpp_dpw_subschemes}
		\end{figure}

		Depending on the operating modes of the Nodes ($X, Y, Z$), up to two \gls{dpw} \emph{sub-schemes} can be derived from the situation shown in the diagram (Fig. \ref{fig:dp_pos}).
		That is possible since the order of pulse and mirror is not fixed in \gls{dpw} \cite{dpw}.
		The signals of $Z$ could therefore either be seen as a mirror-reply, or as a pulse to (potentially) capturing $Y$ and $X$.
		Vice-versa the signals of $Y$ can be interpreted as pulse or as mirror-reply to $X$ and $Z$.
		Both sub-schemes are illustrated in Fig.~\ref{fig:dpp_dpw_subschemes}. 
		Attention, mind the different ordering of the Nodes compared to the original \gls{dpw} paper.
		
		We assume that in a System, operating in \gls{dpp}-mode, any Node triple $(X, Y, Z)$ of the following form
		generates novel input information for a \gls{dpw} calculation:
		\begin{equation*}
			\begin{array}{ccl}
			\forall X \in (N_P \cup N_B), Y \in (N_A \cup N_B), Z \in N_B:\\
			X \neq Y \land Y \neq Z \land X \neq Z
			\end{array}
		\end{equation*}

		First we define the timespans we would measure with any such triple.
		\begin{align}
			R_X^{YZ} &:= r_X^{Zp} - r_X^{Y1} \nonumber\\
			R_Z^Y &:= r_Z^{Y2} - t_{Zp} \label{eqn:rzy}\\
			D_X^{YZ} &:= r_X^{Y2} - r_X^{Zp} \nonumber\\
			D_Z^Y &:= t_{Zp} - r_Z^{Y1} \label{eqn:dzy}\\ 	 
			\Rightarrow R_X^{YZ} + D_X^{YZ} &= R_Z^Y + D_Z^Y \label{eqn:symmetry}
		\end{align}
		with $p \in \{1, 2\}$ being arbitrary but constant.
		It determines which of the two \emph{mirror-replies} should be used for the \gls{dpw} calculation.
		The choice of the pulse does not make a difference in theory.
		However, it can be used in practice to reduce the error.
		Such a reduction can be done for example by averaging results that are based on different $p$s.
		Such a processing step can reduce the gaussian distributed measurement errors that might affect the \emph{mirror-replies'} timestamps.

		Note that all lines except \eqref{eqn:rzy} and \eqref{eqn:dzy} contain measurement input from all three Nodes.
		This shows that they are dependent on them.
		Displaying this relationship motivates the difference in notation.

		We then define compound variables for those measurements: 
		\begin{align}
			\mu_{XZ}^Y 					:&= \frac{R_X^{YZ} R_Z^Y - D_X^{YZ} D_Z^Y}{R_X^{YZ} + D_X^{YZ}} \label{eqn:mu_from_x}\\
						 					&= \frac{R_X^{YZ} R_Z^Y - D_X^{YZ} D_Z^Y}{R_Z^Y + D_Z^Y} \label{eqn:mu_from_z}\\
			\Rightarrow \text{through \eqref{eqn:symmetry} }&\nonumber\\
			\mu_{XZ}^Y 					&= 2 \cdot \frac{R_X^{YZ} R_Z^Y - D_X^{YZ} D_Z^Y}{R_X^{YZ} + R_Z^Y + D_X^{YZ} + D_Z^Y} \label{eqn:mu}
		\end{align}

		Using the findings of \gls{dpw} \cite{dpw} we can use those measurements in combination with knowledge over $d_{XZ}$
		to calculate the \gls{tdoa} value $T_{XZ}^Y$:
		\begin{equation}
			T_{XZ}^Y = \mu_{XZ}^Y - d_{XZ} \label{eqn:dpp_resembles_dpw}
		\end{equation}
    Equation \ref{eqn:dpp_resembles_dpw} can finally be used to calculate the \gls{tdoa} values of \gls{dpp}.

	\subsection{ToA Value}\label{sct:dpp_ds_twr}
		In the following, we derive the formula which can be used to calculate \gls{toa} values based on the values generated by the \gls{dpp} operation model (\ref{sct:operation_dpp}).

\begin{figure}
			\centering
			\includegraphics{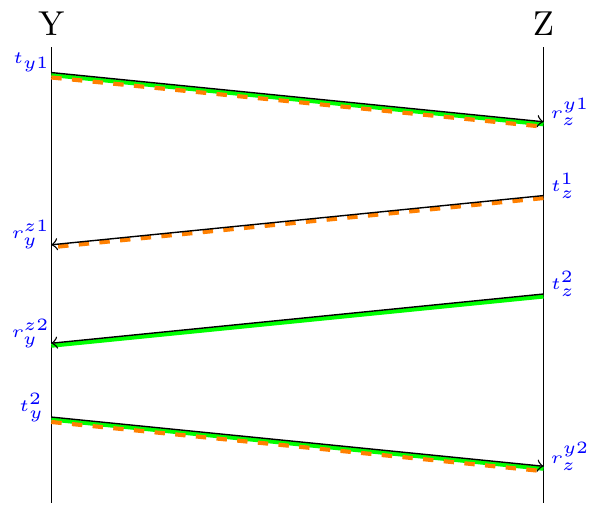}
			\caption{\gls{dpp} extract with two \gls{twr} paths outlined (green(solid) \& orange(dashed))}
			\label{fig:dp_pos_ds}
		\end{figure}

		As we did with \gls{dpw}, we can also identify up to two \gls{asym-ds-twr} schemes in the transmission diagram (Fig. \ref{fig:dp_pos}).
		This becomes evident when inspecting an extract of the diagram and restructuring the visual representation as seen
		in Fig. \ref{fig:dp_pos_ds}.
		When assuming the role distribution $Y,Z \in (N_B)$, we can spot the two possible \gls{asym-ds-twr} measurements we can take.

		Therefore, when performing \gls{dpp} on a System of Nodes, any Node tuple $(Y, Z)$ fulfilling,
		$$\forall Y,Z \in N_B: Y \neq Z$$
		generates new input data for an \gls{asym-ds-twr} calculation (comp. \cite{alternative_twr}).
		Note that the Nodes generating this \gls{twr} input could simultaneously generate the mentioned \gls{dpw} input.

		For such a tuple the timespans measured can be expressed as follows:
		\begin{align}
			R_Y^Z &:= r_Y^{Zq} - t_Y^1 \nonumber\\
			R_Z^Y &= r_Z^{Y2} - t_{Zq}\text{ \eqref{eqn:rzy}} \nonumber\\
			D_Y^Z &:= t_{Y2} - r_Y^{Zq} \nonumber\\
			D_Z^Y &= t_{Zq} - r_Z^{Y1}\text{ \eqref{eqn:dzy}} \nonumber
		\end{align}
		$q \in \{1, 2\}$ is assumed arbitrary but constant.
		It determines which of the possible \gls{ds-twr} paths to use.
		As with the $p$ in the \gls{tdoa} derivation earlier (gaussian distributed) measurement errors, that affect the \gls{ds-twr} paths timestamps, can potentially be reduced.

		Using the findings of \cite{alternative_twr} it follows:
			\begin{align}
				d_{YZ} &:= \frac{R_Y^Z R_Z^Y - D_Y^Z D_Z^Y}{R_Y^Z + D_Y^Z}\label{eqn:ds_dist_from_y}\\
				\text{and }d_{YZ} &:= \frac{R_Y^Z R_Z^Y - D_Y^Z D_Z^Y}{R_Z^Y + D_Z^Y}\label{eqn:ds_dist_from_z}\\
				\Rightarrow d_{YZ} &= \frac{R_Y^Z R_Z^Y - D_Y^Z D_Z^Y}{R_Y^Z + R_Z^Y + D_Y^Z + D_Z^Y}\label{eqn:dpp_ds_dist}
			\end{align}
		Equation \ref{eqn:dpp_ds_dist} is the final formula to calculate the distance between node $Y$ and $Z$.
    This means we can calculate the \gls{tof} between two Bilateral Nodes $Y$ and $Z$ using only the measurements
		gathered in the regular \gls{dpp} operation cycle (\ref{sct:operation_dpp}).

\section{Error Analysis}\label{sct:dpp_error_analysis}
	In the following, we analyze \gls{dpp} for its conditioning towards errors.

	It is well known that clocks, or more precise the oscillators they are based on, in general contain imperfections \cite{allan1987time}. 
	That means that not even
	the highest quality clocks can be trusted with perfect timekeeping. Since \gls{tof}-based
	methods rely on measurements of signals propagating with very high speeds such as the
	speed of sound, or the speed of light $c$, these imperfections pose a major problem regarding
	the accuracy of the localization.
	
	As mentioned in Sec.~\ref{sct:formal_definitions} we assume a clock drift of $\pm20$ ppm which is a typical value for crystal oscillators in the industry \cite{jiang2007asymmetric}.
	We model the clock drift by defining a worst case drift value for every device ($X$) clock separately: $\epsilon_X = \frac{20}{1000000}$.

	\subsection{TDoA Value Error}\label{sct:dpp_tdoa_error}
		Since we deduced all our \gls{tdoa} values $T_{XZ}^Y$ based on the exact method described in \gls{dpw} \cite{dpw},
		we can conduct the error estimation in a very similar way.

		$$\forall X,Y,Z: X \in (N_P \cup N_B), Y \in (N_A \cup N_B), Z \in N_B:$$
		First, the erroneous values are defined:
		\begin{align*}
			k_X &:= (1 + \epsilon_X) \\
			k_Z &:= (1 + \epsilon_Z) \\
			\hat{\mu}_{\epsilon_X} &:= \frac{k_Z \cdot k_X}{k_X}\cdot\frac{R_X^{YZ} R_Z^Y - D_X^{YZ} D_Z^Y}{R_X^{YZ} + D_X^{YZ}} \\
			&\text{and} \\
			\hat{\mu}_{\epsilon_Z} &:= \frac{k_X \cdot k_Z}{k_Z}\cdot\frac{R_X^{YZ} R_Z^Y - D_X^{YZ} D_Z^Y}{R_Z^Y + D_Z^Y} \\
		\end{align*}
      Then using \eqref{eqn:symmetry}, \eqref{eqn:mu_from_x} and \eqref{eqn:mu_from_z} it follows:
		\begin{align*}
			\hat{\mu}_{XZ}^Y :&= \frac{k_X + k_Z}{2} \cdot \mu_{XZ}^Y \\
			\Rightarrow \hat{\mu}_{XZ}^Y &= (1 + \frac{\epsilon_X + \epsilon_Z}{2}) \cdot \mu_{XZ}^Y
		\end{align*}
		Where $\hat{\mu}_{\epsilon_X}$ and $\hat{\mu}_{\epsilon_Z}$ designate the error with the calculation method \eqref{eqn:mu_from_x} or \eqref{eqn:mu_from_z}.
		While $\hat{\mu}_{XY}^Z$ represents the erroneous $\mu$ value resulting with application of the combining transformation from \eqref{eqn:mu}.

		The overall worst-case error in the Node combination $\tilde{\mu}_{XZ}^Y$ is then the difference between the true and erroneous value:
		\begin{align}
			\tilde{\mu}_{XZ}^Y	&:= (1 + \frac{\epsilon_X + \epsilon_Z}{2}) \cdot \mu_{XZ}^Y - \mu_{XZ}^Y\nonumber \\
										&= \frac{\epsilon_X + \epsilon_Z}{2} \cdot \mu_{XZ}^Y \label{eqn:dpp_mu_error}
		\end{align}
		Consequently, the overall \gls{tdoa} error $\tilde{T}_{XZ}^Y$ follows:
		\begin{align*}
			\tilde{T}_{XZ}^Y	&:= \hat{T}_{XZ}^Y - T_{XZ}^Y \\
								 	&= \hat{\mu}_{XZ}^Y - d_{XZ} - \mu_{XZ}^Y + d_{XZ} \\ 
									&= \frac{\epsilon_X + \epsilon_Z}{2} \cdot \mu_{XZ}^Y
		\end{align*}
		We then add and subtract $\frac{\epsilon_X + \epsilon_Z}{2} \cdot d_{XZ}$:
		\begin{align*}
			\tilde{T}_{XZ}^Y &= \frac{\epsilon_X + \epsilon_Z}{2} \cdot \mu_{XZ}^Y - \frac{\epsilon_X + \epsilon_Z}{2} \cdot d_{XZ} + \frac{\epsilon_X + \epsilon_Z}{2} \cdot d_{XZ} \\
			\Rightarrow \tilde{T}_{XZ}^Y &= \frac{\epsilon_X + \epsilon_Z}{2} \cdot T_{XZ}^Y + \frac{\epsilon_X + \epsilon_Z}{2} \cdot d_{XZ}\text{ using \eqref{eqn:dpp_resembles_dpw}} \\
		\end{align*}
		The maximal value $T_{XZ}^Y$ can take on is apparently $d_{XZ}$ (that is the maximal difference of arrival time).
		Therefore we can assume $T_{XZ}^Y \leq d_{XZ}$, from what follows:
		\begin{equation}
			\tilde{T}_{XZ}^Y \leq (\epsilon_X + \epsilon_Z) \cdot d_{XZ} \label{eqn:dpp_tdoa_error}
		\end{equation}
		Which aligns with the error estimations from \gls{dpw} \cite{dpw}.
		That means that the \gls{tdoa} value of the \gls{dpp} method has the same worst case (clockdrift induced) error
		as \gls{dpw}.
		

	\subsection{ToA Value Error}\label{sct:dpp_toa_error}
		We used the findings of the aysm-DS-\gls{twr} to calculate the distance values between two Bilaterals (\ref{sct:dpp_ds_twr}).
		We can therefore apply the same method for its error estimation.
		For all $X,Z \in (N_B): X \neq Z$ the \gls{dpp}-method-term calculating $d_{YZ}$ \eqref{eqn:dpp_ds_dist} is besides a renaming ($X \rightarrow A, Z \rightarrow B$)
		equivalent to the one described in the works of \citet{alternative_twr}.
		Therefore we can apply the same transformations to \eqref{eqn:dpp_ds_dist}, which results in the error estimate:
		\begin{equation}
			\tilde{d}_{XZ} = \frac{\epsilon_X + \epsilon_Z}{2} \cdot d_{XZ}
		\end{equation}
		Therefore the \gls{toa} (clock drift induced) error of \gls{dpp} is equivalent to the one of \gls{asym-ds-twr}


	\subsection{Error Analysis Summary}
		We can, therefore, estimate that the error each measurement can be characterized
		as:
		\begin{equation*}
			\begin{array}{ccl}
				\forall X,Y,Z: X \in (N_P \cup N_B), Y \in (N_A \cup N_B), Z \in N_B:\\
				\tilde{T}_{XZ}^Y \leq (\epsilon_X + \epsilon_Z) \cdot d_{XZ}$$
			\end{array}
		\end{equation*}
		and 
		$$\forall X,Z: X,Z \in N_B: \tilde{d}_{XZ} = \frac{(\epsilon_X + \epsilon_Z)}{2} \cdot d_{XZ}$$

		In comparison with the worst-case error estimated of compared methods \cite{dpw,alternative_twr,hach_ss_twr,Whistle,anti_whistle}
		the proposed method is always at least as well (error) conditioned
		as the compared ones.

		The detailed and unified analysis and comparision of the errors of \cite{dpw,alternative_twr,hach_ss_twr,Whistle,anti_whistle}
		can be found in our other publication \cite{analysis_paper}.

		Finally, it is noted that the equality stated in \eqref{eqn:xyz_eq_zyx} does not necessarily translate
		for its errors.
		In other words, $\tilde{\mu}_{YZ}^X$ does not need to be equal to $\tilde{\mu}_{XZ}^Y$
		as they depend on different clocks.
		One depends on $\frac{\epsilon_Y + \epsilon_Z}{2}$ and the other on $\frac{\epsilon_X + \epsilon_Z}{2}$.

\section{Evaluation of DPP}\label{sct:dpp_advantages}
	In the following, we analyze how \gls{dpp} is related to other \gls{tof} method in the literature and how its \gls{tdoa} and \gls{toa} calculation is improving
  existing methods..

\subsection{Relation to other \gls{tdoa} methods}
We can then extend and re-arrange \eqref{eqn:dpp_resembles_dpw} with \eqref{eqn:dpp_tdoa_definition}:
\begin{align}
  \Leftrightarrow d_{YZ} - d_{XY} &= \mu_{XZ}^Y - d_{XZ}\nonumber \\ 
  \Leftrightarrow d_{XZ} - d_{XY} &= \mu_{XZ}^Y - d_{YZ}\nonumber \\ 
  \Leftrightarrow T_{YZ}^X &= \mu_{XZ}^Y - d_{YZ} \label{eqn:dpp_resembles_djkm}\\ 
  \Leftrightarrow d_{YZ} + d_{XZ} &= \mu_{XZ}^Y + d_{XY} \nonumber
\end{align}
Those equations can be interpreted as follows: With knowledge of one distance between either $X, Y$ or $Y, Z$
we can use the measurement $\mu_{XZ}^Y$ to calculate the \gls{tdoa} value $T_{XZ}^Y$ respectively $T_{YZ}^X$.

If we compare the roles of the Nodes in the setup ($X \in (N_P \cup N_B), Y \in (N_A \cup N_B)$ and $Z \in N_B$) we can make two observations:

\paragraph{\gls{djkm} can be derived}\label{sct:djkm_can_be_derived}
  We look at the case $X \in N_P, Y \in N_B$, and $Z \in N_B$, rename $X \rightarrow T, Y \rightarrow AN0, Z \rightarrow AN1$ and
  remove the secondary pulses. 
  Renaming is valid w.l.o.g. here as the capabilities of the Nodes we rename to include the capabilities we rename from.

  Now the setup exactly resembles the \gls{djkm} method \cite{anti_whistle}.
  In addition, our complete \gls{dpp} setup calculates the same \gls{tdoa} information as \gls{djkm}
  (comp. \eqref{eqn:dpp_resembles_djkm}) would.

\paragraph{\gls{dpw} can be derived}
  As already mentioned, \gls{dpw} schemes can be derived from the \gls{dpp} scheme.
  In the case $X \in N_P, Y \in N_A$, and $Z \in N_B$ this is achieved by renaming $X \rightarrow A, Y \rightarrow S, Z \rightarrow B$
  and removing the second pulse from $Z$.
  Then this sub-scheme exactly resembles \gls{dpw} (comp. \cite{dpw}).
  Further, we calculate with \gls{dpp} the exact same \gls{tdoa} information (compare \eqref{eqn:dpp_resembles_dpw}) as the \gls{dpw}
  that was derived would.

For the particular case $X \in N_B, Z \in N_B$ we can also deduce:
\begin{align} 
  \mu_{XZ}^Y &= d_{YZ} - d_{XY} + d_{XZ} \nonumber\\
  \mu_{ZX}^Y &= d_{XY} - d_{YZ} + d_{XZ} \nonumber\\ 
  \Rightarrow \mu_{XZ}^Y + \mu_{ZX}^Y &= 2d_{XZ} \label{eqn:mu_magic1}\\
  \Rightarrow d_{XZ} &= \frac{\mu_{XZ}^Y + \mu_{ZX}^Y}{2} \label{eqn:bbx_direct}
\end{align}
This value seems to be neither of the type \gls{toa} nor \gls{tdoa} and is therefore of particular interest for 
acquiring a deeper understanding of the relation between \gls{toa} and \gls{tdoa}.
Nonetheless an analysis of the properties of this value would be out of the scope of this work and is subject to future work. 

For the special case $X,Y,Z \in N_B$ it can be shown that:
\begin{align}
  T_{XZ}^Y = d_{ZY} - d_{XY} &= \mu_{XZ}^Y - d_{XZ} \nonumber\\
  T_{YZ}^X = d_{XZ} - d_{XY} &= \mu_{YZ}^X - d_{YZ} \nonumber\\
  \Rightarrow \mu_{XZ}^Y &= \mu_{YZ}^X \label{eqn:xyz_eq_zyx}
\end{align}
This follows directly from \eqref{eqn:symmetry} and shows the symmetry between the two values in \eqref{eqn:xyz_eq_zyx}

	\subsection{Accuracy Improvement}\label{sct:dpp_advantage_accuracy}
		For situations in which we try to position a Passive Node the only known applicable method until now was \gls{djkm} (inv-Whistle).
		With \gls{dpp}, there is now a new method for such situations.
		\gls{djkm} has in those cases a worst case \gls{tdoa} error of:
		\begin{equation}
			\approx D_B(\epsilon_T - \epsilon_B)
		\end{equation}
		The worst case \gls{tdoa} error of \gls{dpp} was previously shown (in Section \ref{sct:dpp_tdoa_error}) to be:
		\begin{equation}
			\approx (\epsilon_X + \epsilon_Z) \cdot d_{XZ}
		\end{equation}

		$D_B$ is typically in the order of magnitude of milliseconds.
		It is usually many orders of magnitudes larger than $d_{XZ}$, which is typically in the order of magnitudes of nanoseconds.
		Consequently, the accuracy of \gls{dpp} is orders
		of magnitudes higher.
		This improvement can be attributed to the additional second pulse of \gls{dpp} compared to \gls{djkm}.
		In situations where \gls{dpw} or \gls{twr} are applicable, we have shown that \gls{dpp} has the same worst-case error estimates (\ref{sct:dpp_tdoa_error} and \ref{sct:dpp_toa_error}).
	
	\subsection{Efficient Mobile Positioning}
		\gls{dpp} solves the problem of mobile positioning by performing \gls{toa} and \gls{tdoa} simultaneously.
		That way we only require a minimum of three Bilateral Nodes in our System to position the whole System.
		In practice, we recommend the use of four Bilateral Nodes for error correction.
		
		Figures Fig. \ref{fig:example_pa_bbb_mobile} and Fig. \ref{fig:example_pa_bbbb_mobile} show those two minimalistic examples of the new class of positioning situations
		that becomes available now.
		The positioning is achieved by first using the \gls{toa} values between the Bilateral Nodes to establish their relative positions and
		then using that information for \gls{tdoa}-based positioning of the remaining Nodes.

	\subsection{New Cases}\label{sct:dpp_new_cases}
		\gls{dpp} further allows addressing some situations that could not be positioned by one existing method alone.
		We present some example of such setups:

		\subsubsection{More Flexible Positioning of a Passive Node}
		Until \gls{dpp} the only method which could be used to position a Passive Node was \gls{djkm}.
		To generate \gls{tdoa}-values for a Passive Node, \gls{djkm} requires an infrastructure of Bilateral Nodes \cite{anti_whistle}.
		\gls{dpp} allows in such cases for other, more flexible device roles in the infrastructure.

\begin{figure}
			\centering
			\subfloat[Positioning one Passive Node using three Bilaterals (fixed)\label{fig:example_p_bbb}]
			{
				\includegraphics[width=.4\linewidth]{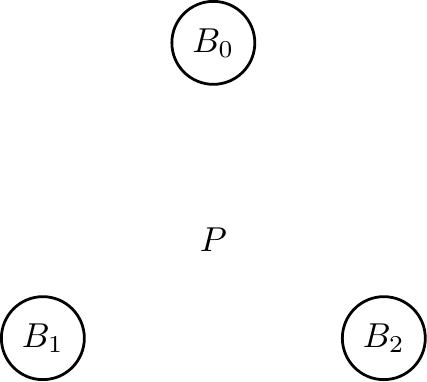}
			}
			\hfill
			\subfloat[Positioning one Passive Node using two Bilaterals and one Active (fixed)\label{fig:example_p_bba}]
			{
				\includegraphics[width=.4\linewidth]{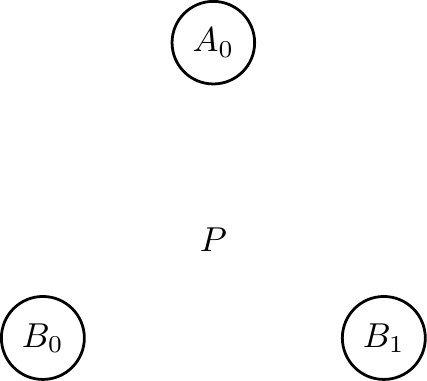}
			}
			\\
			\subfloat[Positioning one Passive Node using three Active and one Bilateral (fixed)\label{fig:example_p_baaa}]
			{
				\includegraphics[width=.4\linewidth]{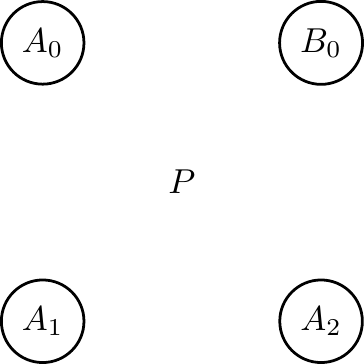}
			}
			\hfill
			\subfloat[Positioning one Active Node using three Passive and one Bilateral (fixed)\label{fig:example_a_bppp}]
			{
				\includegraphics[width=.4\linewidth]{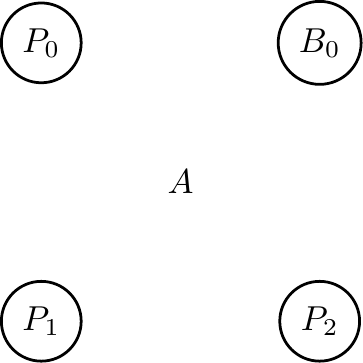}
			}
			\\
			\subfloat[Positioning one Passive Node using four Bilaterals (fixed)\label{fig:example_p_bbbb}]
			{
				\includegraphics[width=.4\linewidth]{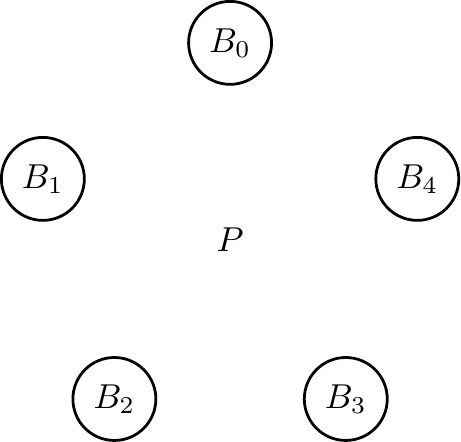}
			}
			\hfill
			\subfloat[Positioning one Passive Node using four Active and one Bilateral (fixed)\label{fig:example_p_baaaa}]
			{
				\includegraphics[width=.4\linewidth]{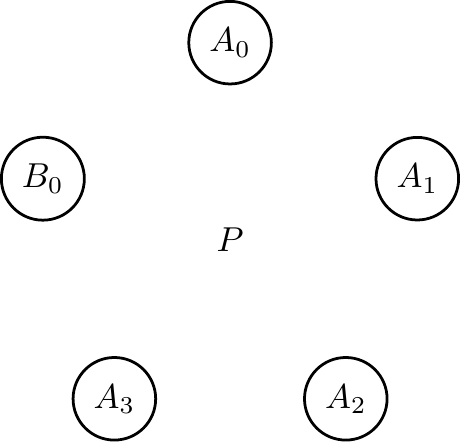}
			}
			\\
			\subfloat[Positioning one (or more) Active or Passive Node using four mobile Bilaterals\label{fig:example_pa_bbbb_mobile}]
			{
				\includegraphics[width=.4\linewidth]{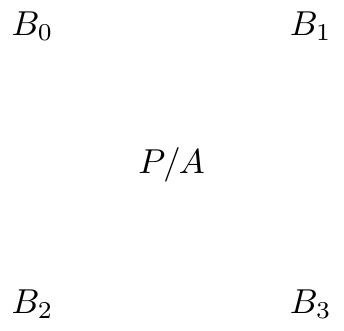}
			}
			\hfill
			\subfloat[Positioning one (or more) Active or Passive Node using three mobile Bilaterals\label{fig:example_pa_bbb_mobile}]
			{
				\includegraphics[width=.4\linewidth]{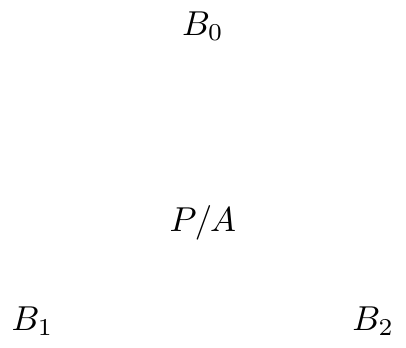}
			}
			\caption{Example Node Systems demonstrating the advantages of DPP}
			\label{fig:example_Node_setups}
		\end{figure}

		For example the minimal setup to create three individual \gls{tdoa} values for a Passive Node using \gls{djkm}
		is depicted in Fig. \ref{fig:example_p_bbb}.
		For \gls{dpp}, the setups in \cref{fig:example_p_bbb,fig:example_p_bba,fig:example_pa_bbb_mobile,fig:example_p_baaa} are all sufficient constellations.

		Concerning the \gls{tdoa} solution engine, for practical reasons it is often even required to generate four \gls{tdoa} values in order to position a Node sufficiently.
		Also for this requirement, \gls{dpp} is an improvement.
		While \gls{dpp} can be used to calculate four \gls{tdoa} values in \cref{fig:example_p_bbbb,fig:example_pa_bbbb_mobile,fig:example_p_baaaa},
		\gls{djkm} will only work for the first case.

		\subsubsection{Mixed Tag Roles}\label{sct:mixed_tags}
			All of the earlier presented known methods are limited to the one type (Active, Passive, Bilateral) of \emph{Tag} they are positioning.
			With \gls{dpp}, this restriction does not apply anymore.
			\gls{dpp} allows for the combination of Active, Passive, and consequently Bilateral \emph{Tags} to be positioned in the same operation cycle.

		\subsubsection{Gathering of TDoA and ToA values}
			Next to \gls{djkm}, \gls{dpp} is the only method known to us, which generates \gls{toa} and \gls{tdoa} values simultaneously.
			Moreover we have shown that \gls{djkm} has a worst-case error larger than \gls{dpp}. 
		
	\subsection{Improved Performance}
		Another benefit which comes with the generalization of \gls{tdoa} and \gls{toa} methods into \gls{dpp} is a reduction in redundancy.
		\gls{dpp} uses in almost all cases equal or very often even fewer messages to gather the same set of measurements as the existing methods.

		\subsubsection{Improvement Compared to DPW}
			\gls{dpw} requires to send three messages for every combination of \emph{Tags} and \emph{Mirrors} in the System \cite{dpw}.
			For a given amount of \emph{Mirrors} $m$ and \emph{Tags} $t$ \gls{dpw} needs at least $n_{DPW}$ messages for a complete operation cycle, where $n_{DPW}$:
			$$n_{DPW} = 3mt$$
			For instance, \gls{dpw} needs $3\cdot1\cdot1=3$ signals when operating on the System displayed in Fig.~\ref{fig:example_a_bppp}.

			\gls{dpp} requires for its operation to send two signals from every Active or Bilateral Node in the System.
			In comparison to \gls{dpw}, \gls{dpp} does require $2\cdot2=4$ signals to operate one complete cycle on the same setup in Fig.~\ref{fig:example_a_bppp}.

			In general, the amount of signals required by \gls{dpp} ($n_{DPP}$) for a system with $t$ Actives (``Tags'') and $m$ Bilaterals (``Mirrors'') is:
			$$n_{DPP} = 2(m+t) \label{eqn:needed_signals_dpp}$$
			Even if we set ``mirror-count'' $m=1$ which resembles the situation in \gls{dpw}, it is apparent that with an increasing number of Tags, \gls{dpp} requires less signals to locate the same 
			amount of Tags, due to the fact that 
			$$2(m+t) \leq 3mt$$
			While the output of \gls{tdoa} values generated are the same for both methods, \gls{dpp} simultaneously yields \gls{toa} values and the new value type from \eqref{eqn:bbx_direct}. 

			It can be assumed that the situation $t \leq 2$ (or even $m = 1$) rarely occurs in real-world scenarios.
			Therefore in almost all realistic Node setups, the application of the \gls{dpp} scheme improves a setup that was previously operated on the \gls{dpw} scheme in terms of sent Nodes/signals.

		\subsubsection{Performance Improvement Against DJKM}
			When comparing \gls{dpp} with \gls{djkm}, one has to bear the mentioned difference in accuracy (Sec.~\ref{sct:dpp_advantage_accuracy}) in mind.
			However, even leaving this advantage aside, \gls{dpp} outperforms \gls{djkm}.

			To substantiate that statement, one has to realize the operation principle of \gls{djkm} as it is defined in the original paper \cite{anti_whistle}.
			\gls{djkm} specifies a distinct order after which Anchors should communicate with each other.
			This coordination is necessary to avoid collisions on the transmission medium.
			This specific order consists of two rounds.
			To describe the ordering, we consider our set of Anchors as an ordered list.

			\paragraph{First Round}
				In the first round,  as described in \cite{anti_whistle} each Anchor performs communication with its successor in the list.
				This process starts with the first Anchor in the list and ends with the last one.
				For example, the first Anchor interacts with the second, then the second interacts with the third, and so on.
			
			\paragraph{Second Round}
				In the second round, only Anchors with an uneven index in the list communicate in the same sequential order with each other.
				Again, this starts with the first Anchor and ends with the last uneven Anchor.
				For example the first interacts with the third, then the third with the fifth, and so on.

\begin{figure}
				\centering
				\includegraphics{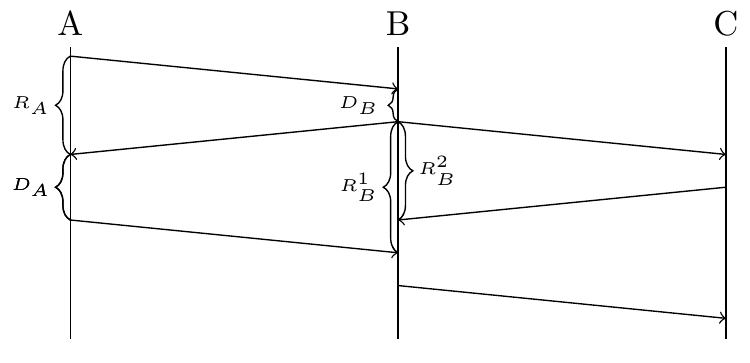}
				\caption{\gls{djkm} with Anchor message re-use}
				\label{fig:reuse_djkm}
			\end{figure}

			This sequential chain of communicating Anchors can be reduced as shown in Fig.~\ref{fig:reuse_djkm}.

			In this Figure, the first message emitted by $B$ is used to initialize the \gls{djkm} communication with the new Anchor device $C$.
			From $A$s perspective, this signal is still a reply in $A$ and $B$'s  \gls{djkm} communication.
			For $C$ this signal resembles the initial message in $B$ and $C$'s \gls{djkm} communication.
			That way, the exchange of one signal can be saved whenever an Anchor has a predecessor since we can then reuse its reply message to the predecessor as the initial message for its successor.
			Note that the first Anchor in the list still has to transmit its initial message. 

			The number of signals ($n$) transmitted in one operation Cycle can therefore be calculated depending on the number of Anchors ($k$):
			\begin{align}
				\forall k > 2:&\nonumber \\
				n_1	&:= (2k - 1) \label{eqn:djkm_signals_transmitted_n1} \\
				n_2	&:= (2\lceil\frac{k}{2}\rceil - 1) \label{eqn:djkm_signals_transmitted_n2} \\
				n		&:= n_1 + n_2 = 2k + 2\lceil\frac{k}{2}\rceil - 2 \label{eqn:djkm_signals_transmitted}
			\end{align}
			Where $n_1$ and $n_2$ are the numbers of signals transmitted in the first and second round.

			The \gls{djkm} System which uses the least amount of signals for gathering three \gls{tdoa}s is shown in Fig. \ref{fig:example_p_bbb}.
			According to the formula derived, the amount of packages transmitted in this scheme is:
			\begin{align*}
				n_1	&= 2\cdot3 - 1 = 5 \text{ \eqref{eqn:djkm_signals_transmitted_n1}}\\ 
				n_2	&= 2\lceil\frac{3}{2}\rceil - 1 = 3 \text{ \eqref{eqn:djkm_signals_transmitted_n2}}\\
				\Rightarrow n		&= n_1 + n_2 = 8
			\end{align*}
			When using \gls{dpp}, according to \eqref{eqn:needed_signals_dpp}  this exact setup takes only $2\cdot3 = 6$ signals to finish the \gls{dpp} cycle.
			According to \ref{sct:djkm_can_be_derived}, it produces the exact same \gls{tdoa} values. 

			In general, the signal requirements of every setup can be optimized by switching the scheme of operation from \gls{djkm} to \gls{dpp}.

			That can be argued as follows.
			A System of $k$ Anchors/Bilaterals and $t$ Tags/Passives operating on \gls{djkm} requires
			$2k + 2\lceil\frac{k}{2}\rceil - 2$ \eqref{eqn:djkm_signals_transmitted} signals.
			\gls{dpp} requires for the exact same setup only $n_{DPP} = 2k$ signals. See \eqref{eqn:needed_signals_dpp}.

			Furthermore, \gls{dpp} allows us in many situations to use Active instead of Bilateral Nodes while achieving the same result as \gls{djkm} would with all Nodes being bilateral.
			For instance \ref{fig:example_p_bba} will yield in one \gls{dpp} cycle the $\mu$ values $\mu_{PB_0}^{A_0}, \mu_{PB_1}^{A_0}, \mu_{PB_0}^{B_1}$ and additionally $\mu_{B_0B_1}^{A_0}$.
			Applying \eqref{eqn:dpp_resembles_dpw} and \eqref{eqn:dpp_tdoa_definition}, the resulting equation-set,
			\begin{align*}
				d_{B_{0}A_0} - d_{PA_0} &= \mu_{PB_0}^{A_0} - d_{PB_0}\\
				d_{B_{1}A_0} - d_{PA_0} &= \mu_{PB_1}^{A_0} - d_{PB_1}\\
				d_{B_{0}B_1} - d_{PB_1} &= \mu_{PB_0}^{B_1} - d_{PB_0}
			\end{align*}
			can then be transformed to 
			\begin{align*}
				d_{PB_0} - d_{PA_0} &= \mu_{PB_0}^{A_0} - d_{B_{0}A_0}\\
				d_{PB_1} - d_{PA_0} &= \mu_{PB_1}^{A_0} - d_{B_{1}A_0}\\
				d_{PB_0} - d_{PB_1} &= \mu_{PB_0}^{B_1} - d_{B_{0}B_1}
			\end{align*}
Under knowledge of the circled Nodes positions, this gives the \gls{tdoa} values $T_{A_0B_0}^P, T_{A_0B_1}^P$ and $T_{B_1B_0}^P$.
			\gls{djkm} would result in Fig. \ref{fig:example_p_bbb}, a fully bilateral infrastructure with the measurement set $T_{B_0B_1}^P, T_{B_1B_2}^P, T_{B_0B_2}^{B_1}$, which is isomorphic.


			The shown relaxation of restrictions does not only simplify the construction of positioning infrastructure. It also reduces the energy consumption of the whole network.
			That reduction roots in the fact that the energy consumption of radios that are commonly used for \gls{tof} is in practice orders of magnitudes larger when operating in RX mode rather than in TX \cite{dw_power_consumption}.
			
		\subsubsection{Bilaterals}
			Bilateral Nodes can be interpreted as operating active and passive at the same time.
			It was already mentioned in Sec.~\ref{sct:mixed_tags} that both types of Tags get processed in the same cycle when using \gls{dpp}.
			This way, the preceding two paragraphs both apply simultaneously to any Tag with type Bilateral, generating the discussed \gls{tdoa} values. 
			That information can be used as input for a \gls{tdoa} solver. 
			
			At the same time we already discussed that any pair of Bilaterals yields a \gls{toa} value through \gls{asym-ds-twr} when performing \gls{dpp}.
			Again, this measurement comes at no additional performance cost as the same signals get used for \gls{twr}, \gls{dpw}, and direct values at the same time.
			While the authors of \gls{djkm} mention that their System does measure \gls{toa} values between the Anchors, this does only work for certain pairs of Anchors \cite{anti_whistle}.
			It also does not work for Tag-Nodes with the role active or bilateral.


\section{Conclusion and Future Work}\label{sct:comparision}
	In this work we presented a novel \gls{tof} method, capable of tracking the relative positions of a large number of devices without the need for a fixed infrastructure.
	It is capable of collecting \gls{tdoa} and \gls{toa} values simultaneously (\gls{tdoa}: \eqref{eqn:dpp_resembles_dpw} and \gls{toa}: \eqref{eqn:dpp_ds_dist}).

	We have proven our method to be more flexible, accurate, and efficient than existing methods.
	
	Our \gls{tof} positioning is therefore suitable for applications such as swarm drone coordination, Indoor Navigation/Guidance, and autonomous driving.


Since we excluded the side channel communication between the nodes from our analysis, future work has to include the protocol overhead for coordination of the pulses 
of Active Nodes for a realistic implementation of the proposed method.


\bibliography{dpp}
%

\end{document}